# ICTD for Healthcare in Ghana: Two Parallel Case Studies

Rowena Luk, Matei Zaharia, Melissa Ho, Brian Levine, and Paul M. Aoki

*Abstract* — This paper examines two parallel case studies to promote remote medical consultation in Ghana. These projects, initiated independently by different researchers in different organizations, both deployed ICT solutions in the same medical community in the same year. The Ghana Consultation Network currently has over 125 users running a Web-based application over a delay-tolerant network of servers. OneTouch MedicareLine is currently providing 1700 doctors in Ghana with free mobile phone calls and text messages to other members of the medical community. We present the consequences of (1) the institutional context and identity of the investigators, as well as specific decisions made with respect to (2) partnerships formed, (3) perceptions of technological infrastructure, and (4) high-level design decisions. In concluding, we discuss lessons learned and high-level implications for future ICTD research agendas.

## I. INTRODUCTION

ICTD research focuses primarily on matters relating to its target populations and their conditions, such as how individuals, communities and institutions interact with technology. For the most part, investigators only become part of the scene as objects of retrospective critique, usually in the context of having failed to see some vital point and achieve some critical goal. As with any kind of research, however, ICTD research is a product of social, cultural and professional influences on the investigators, influences that affect every decision made over a project's lifetime. It is extremely difficult to reflect upon the effects of these influences because each project is so uniquely framed by the contexts of its investigators and its target setting that it is hard for analysts to imagine how "it could have been otherwise" [19].

In this paper, we address this reflective gap through a side-by-side analysis of two completely independent projects which arose from similar objectives but resulted in two very different strategies. Two sets of researchers came to the same country to work with the same community, identified the same problems, and proposed two different solutions. Both projects were implemented over the course of 2007-2008, both aimed to tackle remote medical consultation among Ghanaian doctors, and both sought nationwide deployment over the course of the project lifetime. However, the Ghana Consultation Network (GCN) was initiated by a group of technologically-oriented researchers, while the OneTouch MedicareLine (ML) was initiated by a public health researcher and social entrepreneur. GCN's solution consisted of a Web application hosted on a delay-tolerant network of computers running in each hospital and on the open internet. ML focused instead on a mobile phone program involving a combination of technological services and business innovation. We discuss the rationale behind decisions made by each party over the course of the project and the resulting outcomes.

The main contribution of this paper is to reflect on some ways in which the context and background of researchers affect the structure of ICTD projects. A second contribution is in identifying relevant factors which will assist in the design and execution of ICTD projects in the future. In particular, we provide specific examples of how an investigator's institutional context and identity affect not only the methodology used but also the interpretation of findings; how the partnerships chosen for co-development and co-deployment have a fundamental role in the development and deployment of the technology; how 'objective' assessments of existing technology infrastructure are influenced by personal areas of expertise; and how projects can be designed with various degrees of technological 'specificity' and the resulting implications for their impact on various development indicators. We expect these findings to be of interest to the community of ICTD researchers and practitioners as a whole.

The paper is organized as follows. We first present a background overview of some of the underlying issues of healthcare in Ghana and briefly introduce both projects. We then discuss in turn the framing of the research problem and how it affected the partnerships formed and resulting deployment strategies; the assessments of technological infrastructure by the project members; and the usage and appropriation of the two solutions. Finally, we discuss high-level implications for ICTD research and present works related to each of these sections.

Manuscript received September 22, 2008. This material is based in part upon work supported by the U.S. National Science Foundation under Grant No. 0326582. M. Zaharia is supported by the Natural Sciences and Engineering Research Council of Canada

R. Luk is with AMITA Telemedicine Inc., Toronto, ON, Canada (e-mail: rowena@amitatelemedicine.org).

M. Zaharia is with the Department of Electrical Engineering and Computer Sciences, University of California, Berkeley, CA, USA (e-mail: matei@berkeley.edu).

M. Ho is with the School of Information, University of California, Berkeley, CA, USA (e-mail: mho@ischool.berkeley.edu).

B. Levine was with New York University, New York, NY. He is now with the Department of Obstetrics and Gynecology, New York Presbyterian Hospital, Columbia University Medical Center, New York, NY, USA. (email: DrBrianLevine@gmail.com).

P.M. Aoki is with Intel Research, Berkeley, CA, USA (e-mail: aoki@acm.org).

## II. BACKGROUND

In this section, we highlight some of the background issues of healthcare in Ghana and provide an overview of the two projects which are the focus of our case studies.

### A. Healthcare in Ghana

Like many other countries, Ghana has a tiered healthcare system in which cases that cannot be handled by an institution at a given tier are referred to institutions above it in the hierarchy. All referral chains culminate in one of two teaching hospitals, both located in the South, where specialists have the training and resources necessary to carry out more complex procedures. Doctors, and in particular specialists in areas such as internal medicine, are highly concentrated in the urban South. This sometimes leaves only two or three doctors to serve in district hospitals in the rural North; predictably, rural doctors are confronted with heavy workloads (meaning they can spend only a few minutes face-to-face with a given patient per day) and isolated working environments (which prevent them from taking advantage of many of the educational and collaborative programs available to doctors in the South).

Formal and informal consultation is highly integrated into the life of all doctors. Between hospitals, many doctors call personal contacts – friends, colleagues, classmates – to seek advice, and within hospitals this behavior is even more frequent. Such "curb-side" consultation has long been a common observed characteristic of medical practice [16] but it is of critical importance in environments where specialist expertise is spread thinly.

Continuity of care is difficult to ensure. Patient records are almost universally paper-based. Despite attempts by some major hospitals to transition towards an electronic system, none of the 18 hospitals visited had a working system. Within a major hospital, there is nothing to guarantee complete patient records besides a doctor's own discretion and multiple reminders from administration. It is not uncommon for records to be lost, confused, or incomplete.

The communication infrastructure in the country is severely limited. Hospital landlines are frequently out of service, forcing many doctors to rely on personally-purchased mobile phones. Likewise, broadband Internet infrastructure is unevenly available, frequently unreliable where available, and dependent on hospital budget allocations. Other options are available throughout the country, such as dial-up, satellite, and mobile data plans, but these are generally expensive, unreliable, and/or slow as well.

### B. The Ghana Consultation Network (GCN)

*Solution*

GCN is an end-to-end, computer-based system providing doctor-to-doctor medical consultation on a network of servers implementing a distributed, asynchronously-synchronized database (Figure 1) [20]. The goal is to allow doctors throughout Ghana to consult with each other as well as medical professionals (in particular, the large Ghanaian medical diaspora) around the world. Doctors access the system through a Web-based UI (Figure 2), either by logging into a local server (hosted at some participating hospitals) or by logging into one of the two public servers (hosted with Internet service providers (ISPs) in both Ghana and the U.S.). Providing local servers ensures availability and responsiveness to the users in the face of unreliable network connectivity and makes the task of synchronizing data between servers transparent to them. Synchronization is automated and carried out over a disconnection-tolerant messaging layer called OCMP [29] which draws inspiration from the research into delay-tolerant networking (DTN) [7],[12].

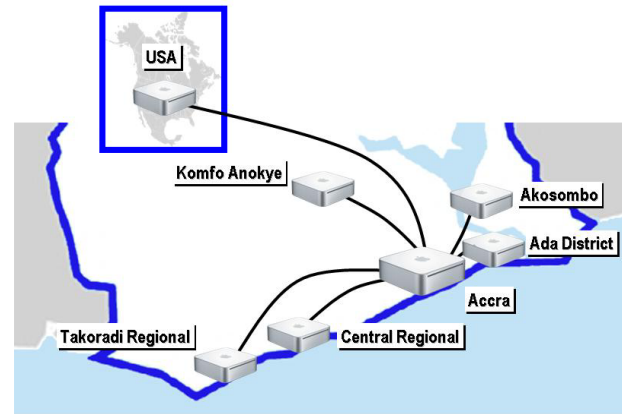

Figure 1. GCN distributed, delay-tolerant server network

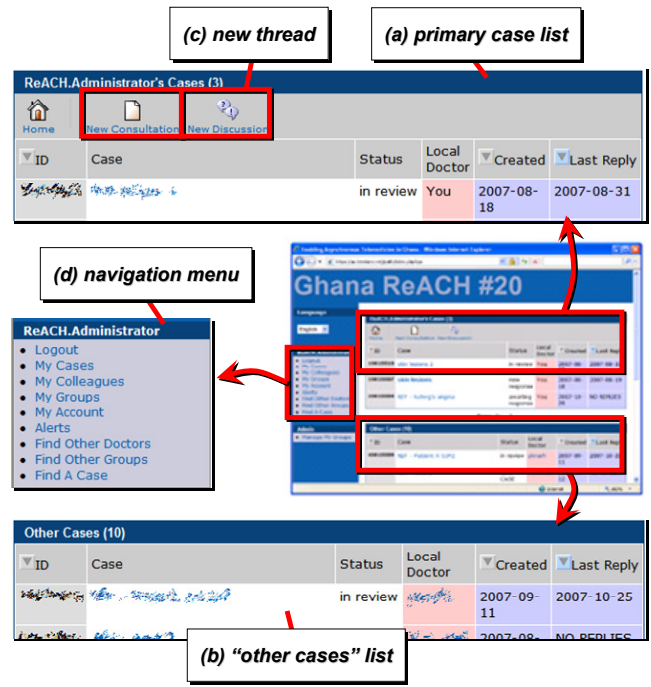

Figure 2. Key elements of GCN's welcome screen

Because the doctors already view consultation as a matter of reaching out to personal contacts, the system is presented as a social networking platform – a forum for medical consultation with social and professional colleagues – and leverages social incentives using principles drawn from the HCI and CSCW communities [20]. The system supports two types of 'conversations': highly structured 'consultations' for specific patients (which work much like an electronic case history) and unstructured 'discussions' (which work much like online forums). So, for example, a doctor unsure of how to treat a particularly resistant case of tuberculosis might fill out various fields of the 'consultation,' address it to a colleague, and wait for advice to be appended to that consultation, whereas another interested in general updates on malaria treatment might create a question under 'discussions,' addressed to any interested doctor. Recognizing the ubiquity of mobile phones, GCN also incorporates text message (as well as email) notifications of new content; yet the core of the interaction is designed for computers. This is based on feedback during the design process that the wealth of data required for patient management would demand a screen size larger than those present in mobile phones.

To date, over 125 doctors have been enrolled from Ghana, the U.S., Mali, Nigeria, South Africa, and the U.K. and 39 consultations have been submitted.

*Methodology*

GCN is the product of a conventional user-centered design process. There have been four rounds of iterative design and fieldwork, starting with exploratory needs assessment (2005) and continuing in conjunction with design exercises (2006-2007), a pilot deployment (2007), and an ongoing deployment (2008). Overall, interviews or focus groups were conducted with 132 doctors in 15 hospitals throughout Ghana [20].

The most recent fieldwork in Ghana lasted six weeks in mid-2008. Exploratory interviews on mobile phones and computer usage were conducted with 35 internal medicine (IM) doctors at a major teaching hospital. Further, six evaluation interviews at one regional hospital and two interviews at one district hospital were conducted (both sites of the original pilot). Limitations of the methodology include an overrepresentation of internal medicine doctors from urban hospitals.

While the initial rounds of fieldwork focused on rural hospitals which are more numerous in the North, the current live deployment is centered in the more accessible South, with plans to extend later. We have established public servers in both Ghana and the US, as well as local servers in 3 major hospitals and 2 rural district hospitals in the South. These district hospitals suffer from some of the same problems as in the North – a shortage of doctors and poor travel and communication infrastructure – although the severity of these challenges is much less. Our current strategy is first to recruit specialists from the major hospitals and to test our system in the district hospitals of the South before reaching out to more challenged environments in the North.

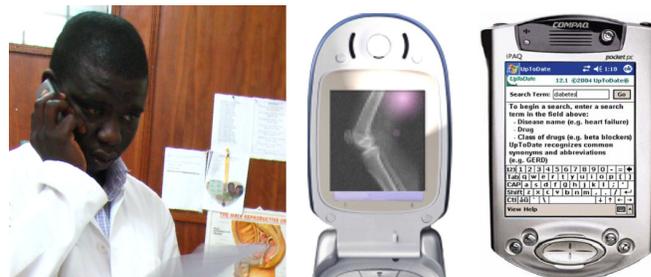

Figure 3. ML Phase 1 calls for free phone calls and text messages; the planned ML Phase 2 calls for MMS and data reports over SMS, and ML Phase 3 calls for free smartphones, reference tools, and custom applications.

### C. The OneTouch MedicareLine (ML)

*Solution*

MedicareLine is a program currently offering free calls and text messages between any registered physician and/or surgeon within Ghana. Its current focus has been on reducing logistical and economic barriers to mobile phone use rather than on technological innovation. After submitting the required paperwork, a doctor registered with the Ghana Medical Association (GMA) receives a OneTouch GSM SIM which can be used with a privately-purchased mobile phone. Using this SIM, the physician can now call other program participants free of personal charge. For example, a physician can call a specialist in the capital or a friend in a rural town to ask or provide medical consultation. This can be a significant cost saving, especially given that airtime in Ghana is relatively expensive compared to many developed countries. The GMA also has a computer terminal that can send "blast" texts to all participants for updates and notifications.

Future phases of the program envision new technological interventions (Figure 3). Phase 2, as yet uncompleted, calls for physicians to receive free MMS service so that they can augment their phone consultations with photos (e.g., of a skin condition or X-ray). Phase 2 will also allow the GMA and other government organizations to collect data from physicians via SMS. In Phase 3, ML anticipates partnering with hardware vendors to provide each physician with a smartphone preloaded with medical reference software.

Phase 1 of this program has already experienced a very high rate of adoption. Approximately 1700 of 2000 doctors in the GMA have enrolled, with over 2 million calls made to date.

*Methodology*

ML arises from a five-week visit to Ghana in October 2007 and a two-week visit in March 2008. The purpose of the first visit was to introduce and promote usage of a social networking Web system for international medical collaboration that had already been developed in the States; however, within two weeks of arriving in Ghana, the investigator decided that access to both computers and the Internet was a fundamental problem. Discarding the Web-based project entirely, he moved on to exploratory interviews with all stakeholders, including over 30 doctors, politicians, local businessmen, and administrators, focusing on the issue of barriers to communication, innovation, and current technology usage. He conducted site visits to five regional hospitals and one polyclinic in the urban South, and visited Tamale, a northern regional capital, for four days. Towards the end of that trip, the investigator orchestrated a "meeting of the minds" between the CEO of OneTouch and the head of the GMA, whereupon an initial understanding was reached. Implementation was left to the two partnering agencies, which announced the program the following month and launched it in January 2008. In March 2008, the investigator returned to conduct another round of unstructured interviews with over 20 individuals, again from diverse communities, in order to assess the program's progress.

Table 1 – Comparison of GCN and ML

|  | **Ghana Consultation Network (GCN)** | **OneTouch MedicareLine (ML)** |
|---|---|---|
| **Primary platform** | Computers | Mobile phones |
| **Adoption** | 89 doctors in 5 hospitals in Ghana<br>>125 doctors total from around the world | >1700 doctors in Ghana |
| **Problem addressed** | Network connectivity | Cost of cellular airtime |
| **Target user** | Doctors in Ghana and Ghanaian medical diaspora | Doctors in Ghana |
| **Partners (see Table 2)** | Ministry of Health, GPSF, and KNet (ISP) | GMA and OneTouch (mobile operator) |
| **Assumptions** | Adequate internet connection quality and coverage punctuated by regular power and network outages | Adequate mobile phone quality and coverage |
| **Specificity (see Section VI)** | High – custom system software and Web application | Low – only generic mobile phone service |
| **Deployment strategy** | Incremental (hospital by hospital) | National (available country-wide at launch) |

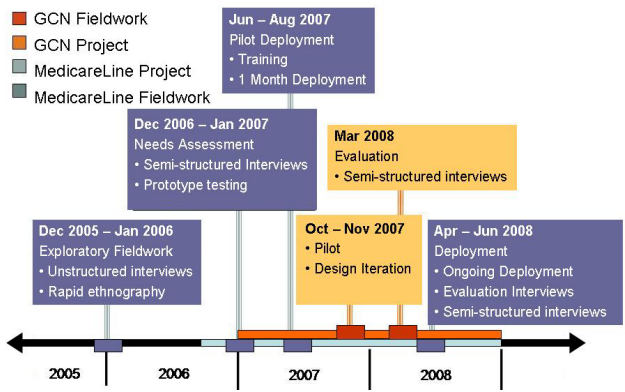

Figure 4. Timeline of GCN and ML Projects, including fieldwork.

In addition to ML's own follow-up interviews, material for this paper also comes from 10 interviews conducted by the GCN investigators with doctors at one of the teaching hospitals, as well as their interviews with the ML investigator directly.

### III. THE ROLE OF PROJECT CONTEXT IN FRAMING PROBLEMS

ICTD work frequently involves the interdisciplinary participation of various communities of research and of practice [5], but what are the implications of working within these communities, given similar projects with similar goals over similar timeframes? Both projects clearly did some amount of fieldwork in both the urban South and the rural North, both purportedly wanted to develop a solution that would work for all Ghanaian doctors, but GCN focused on connectivity in the rural North using innovative technology while ML focused on building communication in the urban South using proven technology. In this section, we examine the institutional, cultural, and personal contexts from which each project arose.

#### A. GCN

GCN began as part of an ongoing collaboration between a U.S. corporate research laboratory and a U.S. research university. Specifically, it was a project of a joint research group which had worked extensively in the area of low-cost connectivity and delay-tolerant networking. Project funding came from U.S. government grants as well as the corporate sponsor. In the first brainstorming/conception phase of this project, the axes by which ideas were evaluated were defined as: (1) direct social impact (e.g., improvements in healthcare delivery), (2) medium-term impact on ICT adoption in developing regions (e.g., finding novel ways to make ICT more relevant in addressing local problems), and (3) long-term contribution towards HCI research. The primary short-term deliverables were software, real deployments, and research papers. As for the individual investigators, all came from a technology and research background, collectively with experience in systems, networks, and HCI.

#### B. ML

ML began as a project funded by a social entrepreneurship program and the international health program of the School of Medicine at a U.S. research university. As part of the international research and education component of a medical program, there were no short-term deliverables. The individual investigator comes from a medical program with prior experience in medical research and a personal interest in consumer technology. His personal goal was to identify a project with sustainable, wide-spread social impact that had the potential to be financially self-sufficient and reproducible in other developing countries.

#### C. Discussion

These differing contexts had far-reaching implications for how the problem was framed. First, coming from a more technological background, GCN investigators were in a better position to attempt technological innovations and, in particular given their knowledge of projects such as [9] and [22], on creative ways of addressing poor connectivity. The ML project's lack of technological expertise limited the possibility of developing new technology. Second, ML, given its ties to a major U.S. hospital, had a natural predisposition to tackle problems internal to a specialist center, whereas GCN had no expertise in hospital administration and hence was more inclined to address the simpler logistics of the rural clinic. Third, ML's agenda for large-scale social change predisposed it towards impacting the greatest number of doctors, who are of course concentrated in the urban South with better mobile phone coverage. GCN chose to focus design efforts on the problems of the North (such as those of the 14 doctors servicing half a million people in Ghana's Upper West region), seeing this as the best use of limited resources. These different problem framings have various implications for the solutions developed and implemented, as we discuss in the following sections.

Table 2 – Healthcare institutions / partners

| Acronym | Name and Description |
|---|---|
| GMA | Ghana Medical Association<br>A voluntary association of 95% of Ghana's doctors, representing their interests nationally |
| MOH | Ministry of Health<br>The national health administration, responsible (through the Ghana Health Service) for all public hospitals |
| GCPS | Ghana College of Physicians and Surgeons<br>Responsible for specialist education and certification |
| GPSF | Ghana Physicians and Surgeons Foundation<br>U.S.-based non-profit organization promoting specialist education in Ghana |

## IV. PARTNERING STRATEGY AND DEPLOYMENT

The question of partner selection is a key one for ICTD deployments, as partners are the usual means by which technology makes it from the laboratory to the field. In this section, we examine the impact of local and non-local partners on technology design and deployment.

### A. GCN

GCN chose partners based on the decision to deploy technology in hospitals to connect their doctors with the Ghanaian diaspora (Figure 5). The Ghana MOH was chosen as a local institutional partner because it had the central authority to allow the servers to be installed in public hospitals. With the initial approval of the MOH, GCN was able to work quickly with the public hospitals, conducting 121 interviews with relatively little trouble alongside iterative, incremental deployment of technology. Of course, the drawback of the centralized approach is that it requires convincing a risk-averse bureaucracy in advance, thereby running the risk of 'over-selling' the solution. In addition to the added effort, another unexpected difficulty of working through the MOH was overloaded communication channels. Advertising GCN to the doctors was a challenge because promotional material needed to work through the same mechanisms through which the MOH communicated, for example, minor procedural changes and optional seminars from pharmaceutical companies. So there were multiple instances when administration would send out notifications of GCN training sessions but only a small fraction of doctors would show up. GCN partnered with the Ghana Physicians and Surgeons Foundation (GPSF), a Ghanaian medical diaspora organization, to recruit medical consultants in the U.S. KNet, a small Ghanaian Internet service provider (ISP), provided local Web hosting.

### B. ML

ML worked with a smaller number of partners, all non-governmental (Figure 6). ML chose the GMA as a local institutional partner because it represented the interests of physicians and its leaders would immediately see the benefit of the project to the physicians as individuals, if not to the healthcare system as a whole. The program is framed as a GMA value-added member service (similar to other member benefits such as professional development seminars and quarterly publications). In contrast to the MOH information channels, doctors took personal initiative to read the GMA emails or check GMA bulletin boards; this lead to widespread awareness of ML. OneTouch was chosen as a technological service provider for political and pragmatic reasons. As the investigator stated:

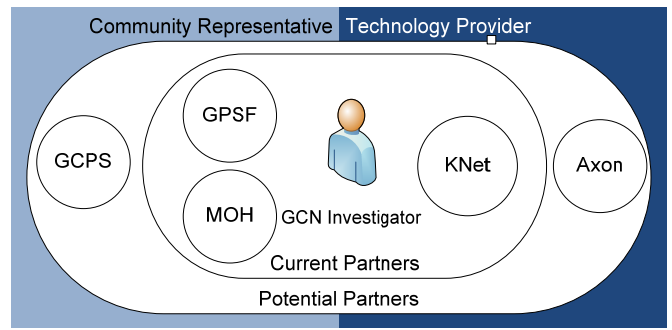

Figure 5. Map of GCN current and anticipated partners.

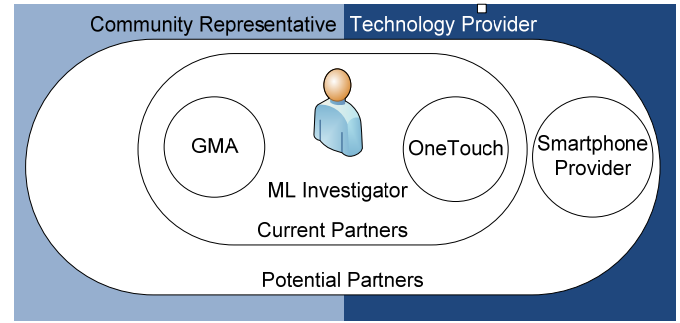

Figure 6. Map of ML current and anticipated partners.

> *OneTouch was the national company. I wanted this to be Ghanaian – by Ghanaians, for Ghanaians. That says a lot more than 'done by Ghanaians in concert with South Africans.' [2]*

OneTouch had the means and the expertise to very quickly make the ML program a reality and the financial resources to support the program independently on an ongoing basis. Thus, while GCN's deployment was incremental, ML's was all-or-nothing, a strategy consistent with ML's social agenda for rapid, wide-scale impact. Of course, the drawback of working entirely through OneTouch was that the project was no longer under the control of the investigator but was now subject to the organizational vagaries of a for-profit corporation. That is, changes in OneTouch's business priorities could result in the program being dropped as quickly as it was initiated. This question is immediately salient in light of the recent acquisition of OneTouch by a multinational carrier based outside of Ghana (in fact, in South Africa).

### C. Discussion

In considering the non-local partners, GCN and ML bear striking similarities in that both were initiated in partnership with U.S.-based organizations hoping to foster communication with the Ghanaian medical community: GCN with GPSF and ML with a U.S. teaching hospital. However, GCN emphasized the role of GPSF throughout the course of its lifetime whereas ML stopped working with U.S. doctors in order to focus explicitly on the Ghanaian context. The investigator noted:

*[B]y removing this component of having this international cross dialogue, and realizing that we needed to have intra-Ghanaian communication instead of inter-Ghanaian communication, I quickly came to the conclusion that I needed to do something to improve communication within Ghana. [2]*

The relevance of overseas medical consultants was ultimately one of the core reasons that ML could focus on mobile phones while GCN retained a focus on the Internet. With respect to deployment, this meant that GCN had the additional task of advertising at GPSF conferences in the U.S. and to other organizations in the West, diluting its efforts in Ghana.

GCN's partnership with the MOH, GPSF and KNet and ML's partnership with the GMA and OneTouch were factors in the projects' very different rates of adoption. Arriving in the country with equipment for the initial pilot deployment, GCN was deployed in four hospitals over the course of five weeks, garnering an initial user base of 73 doctors. Over a similar stretch of five weeks, MedicareLine went from being a conversation between two CEOs to a national program, and by four months later over 1600 doctors had used ML to make over a million phone calls across the country.

On the topic of sustainability, the juxtaposition of these two projects raises some interesting questions surrounding the rhetoric of "**organic adoption**" and its impact on long-term sustainability of ICTD initiatives. That is, they reflect two different views of what 'organic' or 'bottom-up' adoption means.

ML presented a more 'organic' adoption model in the sense that the program was announced and doctors could sign up according to their individual needs and interest. On the other hand, because one institution provides all of the technical and financial resources, there is less local ownership and control over the maintenance of each project. Essentially, there is no guarantee that OneTouch (or the GMA) will not unilaterally end the project. This creates a situation analogous to the many development projects which rely on inconsistent or limited-term donor funding.

GCN's model involves a more laborious adoption process but benefits from complete in-hospital ownership. This is a different type of 'organic'. This 'organic' involved more than just the end user; it involves the whole system of people and machines that need to be in place for this network to grow. Unfortunately, this also surfaces the issue of the lack of access to ICT expertise experienced by all but the largest urban hospitals; the benefits of decentralized ownership are compromised by the geographic concentration of ICT skills.

GCN can be conceived as a centralized project (sponsored by MOH and adopted by hospitals) with decentralized deployment (core resources provided by hospitals) while ML is a decentralized project (adopted independently by individual doctors) with centralized deployment (core resources provided by OneTouch). Technology innovators (such as corporations) often exhibit a bias towards decentralized solutions distributed through markets. Yet there are many examples, of which GCN would be one, of a potentially valuable tool which does not become useful until it has sufficient infrastructure and "network effects" to fulfill its potential. Balancing the costs and benefits of the approaches remains an open question.

V. ASSESSING TECHNOLOGICAL INFRASTRUCTURE

One might think that doing a baseline assessment of available technology infrastructure would be one of the most objective elements of an ICTD project, one that is a basic part of requirements analysis. In this section, we observe that GCN and ML drew on two very different sets of infrastructure assumptions and show that such assessments are highly influenced by personal and institutional context.

*A. GCN*

The GCN researchers designed the system around the pessimistic engineering assumption that the system must continue to function under the worst-case connectivity conditions in the rural North – i.e., that all options available are unreliable and often low-throughput – and optimistic assumptions about PC usage. The exploratory needs assessment fieldwork found that the quality of mobile phone connectivity was unacceptable in the North.

*It is generally so oversubscribed that if you are calling a mobile phone from another service provider, you need to dial 10-15 times in order to successfully connect. [1]*

GCN made no quantitative measurement of the quality and reliability of network connectivity, instead relying on qualitative descriptions of regular power and network outages in formulating design requirements.

With respect to the technology baseline of users, the GCN needs assessment found that:

*About a third of the physicians interviewed access email regularly… whereas another third claimed that they had at some point accessed email regularly. [1]*

Thus, it was expected that only a minority of doctors would actually need to acquire computer literacy.

*B. ML*

The ML researcher made optimistic assumptions about mobile phone connectivity among its user base and pessimistic assumptions about PC usage. As with GCN, connectivity was assessed qualitatively, drawing on experiences during the four-day trip to the North wherein calls from a OneTouch phone were compared with calls from another provider. While there were occasions of calls dropped or text messages delayed, ML concluded that the quality was sufficient for most basic uses.

With regards to computer literacy, ML concluded that:

*The daily routine of a physician in Ghana does not revolve around computers. I can't go a day as a physician [in the U.S.]… without using a computer – taking orders to checking lab values – but in Ghana it's all people ordered, from medical records to orders, operative notes… Everything is done by paper! [2]*

These observations are actually consistent with those of GCN, but what gives these findings an added dimension is the ability of the ML researcher to make a direct comparison of this

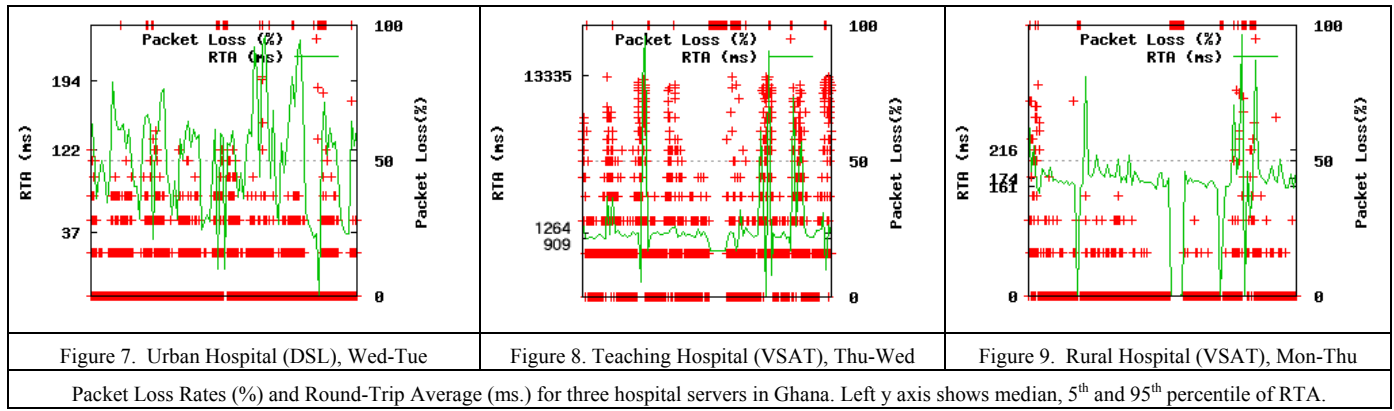

| Figure 7. Urban Hospital (DSL), Wed-Tue | Figure 8. Teaching Hospital (VSAT), Thu-Wed | Figure 9. Rural Hospital (VSAT), Mon-Thu |

Packet Loss Rates (%) and Round-Trip Average (ms.) for three hospital servers in Ghana. Left y axis shows median, 5th and 95th percentile of RTA.

environment with his experience in a leading hospital in the U.S. Thus, while GCN framed its findings as "most doctors do indeed have some experience with computers" and would use them more often given better access, ML concluded that "computers are not part of the daily routine of a Ghanaian physician" and so are not a useful option.

*C. Discussion*

Part of the difficulty in producing consistent assessments of conditions on the ground is that while metrics such as the frequency of power outages and the availability of DSL can be measured and mapped in great detail, it is up to each investigator to determine exactly what metrics need to be measured, how rigorous the measurement needs to be, and what quality of service constitutes something usable by the target community for the specific application. GCN had no resources or expertise with which to improve the mobile phone infrastructure, so it focused more on network infrastructure. ML had no expertise with network infrastructure, but instead saw an opportunity to address a very significant, non-technical barrier in mobile phone usage.

A closer look at the details of connectivity suggests that both projects were somewhat optimistic in downplaying the infrastructural limitations. The qualitative assessments verified expected connectivity barriers, whereas additional issues remained hidden.

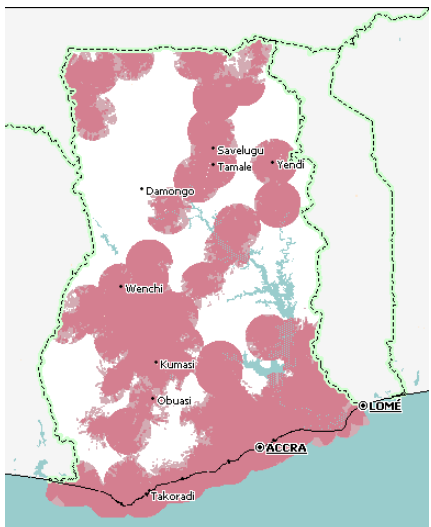

Figure 10. OneTouch GSM Coverage Areas. *(Source: GSM Association)*

GCN anticipated power outages and network disconnections occurring several times a day, but the regularity with which (nominally acceptable) bandwidth was unusably low was a surprise. For example, certain pieces of systems software on the local servers needed to be configured with estimates of the worst-case time needed to transfer an 8KB file chunk; the initial estimate of 20 seconds (~3.3Kb/s) was eventually increased to 5 minutes (~0.2Kb/s). Connectivity data obtained after the deployment of GCN show that the difference between best- and worst-case performance can be extreme. Figures 7, 8 and 9 illustrate the volatility of network performance at three sample hospitals. Figure 7 shows a DSL network connection with moderately variable packet loss rates and round trip times. However, Figure 8 illustrates the case of a satellite connection so overloaded that, on weekdays between 10:00 AM and 4:00 PM, packet round-trip times to the public GCN server in Ghana consistently exceeded 10 seconds and loading google.com took 15 minutes. While the GCN software was successfully able to transmit doctors' messages when congestion decreased in the evening, a doctor who requested a consultation in the morning would wait until the following day to receive a response (even if the consultant responded immediately upon receipt of a message). (The irony here is that this hospital, with the largest number of dedicated IT staff - 6 people - thus experienced the most apparent 'outages'.) In contrast, Figure 9 shows the characteristics of a satellite connection that worked reliably and consistently.

As ML is not based on Internet connectivity, an analogous examination of ML's assessment of mobile phone coverage would require assessing the quality of service of OneTouch voice calls and text messages. While we do not have this information, we can look instead at geographical coverage (Figure 10). While OneTouch has more coverage than any other provider in Ghana, the coverage map shows that vast regions of the country remain out of coverage area, leaving doctors working in those regions at a significant disadvantage. Clearly, a trip to a single urban center in the North does not systematically gauge the limitations of coverage throughout the country, let alone quality of service within current coverage areas.

To be clear, the issues illustrated above have not been a cause of project failure in either case. Further, exhaustive quantitative assessment of all such issues in advance of deployment is not always cost-effective or even possible (for example, one cannot measure DSL links that are not installed). What we suggest here is that care must be taken to prevent "blind spots" in assessing infrastructure and that such blind spots can arise from the investigators' backgrounds.

## VI. Solution Usage and Appropriation

In this section, we discuss the usage of the technological solutions that arose from each project as well as the ways in which the solutions were appropriated for uses other than those they were intended to enable. We then discuss the role of what we call technological specificity in these processes.

### A. GCN

As described previously, GCN adoption has been relatively slow; what is notable is that GCN experienced a better rate of adoption in the smaller hospitals than in the large ones. At the smallest deployment hospital, with only 2 doctors on staff, one of the doctors continues to log into the system and post cases on a roughly bi-weekly basis. In the mid-sized hospitals, the response was mixed, with a handful of doctors using the system every week or two but the majority losing interest. At the largest deployment hospital, however, after the initial flurry of activity surrounding the presentations and training, none of the doctors continued to use it on a consistent basis.

Follow-up interviews shed some light on this disparity. One issue was computer access. In the smallest district hospital, the doctors shared one computer, but since there were only two doctors, having access to the computer was never a problem. In contrast, access to computers was an issue at the larger hospitals. One doctor said:

> *Sometimes you go to the library, you see someone at the computer for 15 minutes; you don't have that time to waste [waiting for the computer].*

Another issue was the match between the system's use case and the needs of the pilot participants. Through its design and pool of GPSF consultants, GCN had targeted general practitioners (GPs) in the North who wished to consult with urban and overseas medical specialists. This matched the needs of district hospital doctors in the South who had no ready access to specialist consultation and saw great value in the ability to connect with specialists (either in Ghana or overseas). The popularity of this system in the district hospitals is a very promising precedent for future deployments in the more rural North. In contrast, physicians at the larger Southern hospitals could speak and consult with other doctors and specialists more easily in-person than over the computer. GCN did not address the needs of urban specialists looking to tap into the global community of sub-specialists. (In evaluation interviews, many of the urban doctors requested a greater number of sub-specialists within the system.)

*GCN Appropriation*

Beyond the expected use for consultation, there were unanticipated uses of the system. First, 6 of the 39 'conversations' observed were purely social in nature. This was surprising in light of how 'medicalized' the investigators perceived the interface. Yet perhaps because their means of communicating with remote doctors are so limited, it appears the barriers between profession and person are much more fluid than anticipated. Second, many of the doctors discussed and requested functionality in the system for sharing literature and PowerPoint presentations, to the extent that these features were included in the upgrade from the pilot to the ongoing deployment. This is consistent with the findings mentioned earlier that a larger screen size was important in dealing with more information-intensive tasks, and also provides insight into the kind of tasks matched to a computer's affordances.

### B. ML

As previously noted, ML experienced an incredible rate of adoption, with 1700 of 2000 members of the GMA signing up within the first four months. While there are no statistics available concerning the fraction of usage that is related to consultation, there is little doubt (judging from the interviews and from multiple instances of observing doctors as they received calls) that ML is used frequently for consultation.

There are multiple reasons for this popularity, some of which are hard to distinguish from the affordances and popularity of mobile phones themselves. (One doctor claimed his phone bill dropped from 150 USD per month to 8 USD per month after joining the program.) First, mobile phones are a popular medium for medical consultation because the real-time nature of voice calls is often critical to treating an emergency case – the three doctors who volunteered information on the breakdown of emergency/non-emergency cases reported that around 80% of cases that require further consultation are indeed emergency cases. One said:

> *I prefer phone calls to SMS, because I prefer an immediate answer, and also so I can make sure the phone is on. If I'm dealing with a case right now, I want to know what to do when moving ahead as soon as possible.*

Second, phone skills are more widespread than computer skills. Because mobile phones are practically ubiquitous in Ghana, there is a lower learning curve as opposed to computers, which are owned by only one in three doctors. A few doctors claimed that texting on a mobile phone was easier than typing on a keyboard. While many of the junior doctors demonstrated great proficiency with both typing and texting – the vast majority sent multiple text messages a day – a few of the senior doctors demonstrated great difficulty typing during training sessions. Third, it promotes more tightly-integrated workflows. One doctor said:

> *I use [ML] a lot and I think it is wonderful. If you want to talk to anybody concerning a case… concerning anything relating to your practice... it gives you a chance to relax and really talk. It's so good. It's a wonderful idea.*

Another said:

> *There has been a move to ban mobile phones in certain hospitals. It is a very very big mistake, because all they are going to realize is that this is actually going to decrease*

*efficiency rather. Consultation is not going to be working as well as it used to.*

*ML Appropriation*

ML was originally framed as a system to ensure continuity of care in long-term and referral patients, but after the system had been established for three months the ML investigator noticed its emergent effect in fostering camaraderie in the medical community. He said:

*I was talking to doctors and they were telling me. "Yeah, I'm reconnecting with classmates."*

Indeed, many doctors were up-front about the fact that ML had been a boon, not just for consultation, but for facilitating social interaction within their community.

The mobile phone has also integrated itself into administrative and management processes within large hospitals. One doctor said:

*You don't have walk down somewhere or you send a patient down... it reduces the whole bureaucratic... pushing around of patients.*

Personal mobile phones, both on the ML program and not, are regularly used to set up diagnostic tests at the laboratory, to confirm insurance claims forms, and also for the doctor on call during the night to discuss a change in a patient's care with the doctor who admitted that patient. In short, where a U.S. hospital might use infrastructure such as pagers and site-wide internal communications systems, large Ghanaian hospitals improvised solutions using personal mobile phones.

*C. Discussion*

In reflecting upon differences between GCN and ML in usage and appropriation, we will focus on an important distinction that we will call 'specificity'. Two solutions with different 'specificity' can be targeted at exactly the same task and be based on an equally nuanced understanding of workflows and use cases; the difference lies in the types and number of layers of technology which make up the solution and the degree to which they are specific to the solution. For example, the GCN system presents not only a robust asynchronous communication medium, but also an in-hospital server as well as an end-user Web application. ML, in its first phase, focuses exclusively on tackling the cost barrier of existing phones using existing networks.

As we have seen, these two projects with similar goals at the start ultimately resulted in two very different usage outcomes – particular in the area of adoption. Investing time and resources in needs assessments and design process, the GCN project produced a highly 'specific' system to address not only failures in internet infrastructure but also social network gaps. However, adoption has been slowed by the need to introduce the system incrementally into hospitals. The ML project achieved broader adoption over a much shorter period of time, in part by relying on the existing availability and high adoption levels of mobile phones. Similarly, the lower specificity of the ML solution seems to contribute to a greater range of user appropriation behaviors.

While it is tempting to conclude that providing solutions with lower specificity is strictly more desirable – and in many cases it may be more desirable – it must also be remembered that utility comes in many forms. GCN's higher degree of specificity is due to multiple factors. First, the social networking application is required to meet the GCN goal of enabling isolated doctors to build social capital in an extended geographic network of colleagues. While ML assumed that the doctors that needed to work together already had each other's mobile phone numbers, GCN had determined that a large portion of rural, immigrant, and junior doctors did not have a strong network of contacts [20]. Second, GCN's emphasis on overseas consultants implies a need for low-bandwidth, asynchronous communication as opposed to voice calls. Third, GCN operated as a development project, maintaining statistics of usage metrics in order to facilitate evaluation; for now, ML relies entirely on the built-in reporting mechanisms in the OneTouch network. GCN can map consultations made to specific case outcomes, while OneTouch – which does receive very high praise from enrolled doctors who offered their feedback on the system – cannot. Moreover, as previously mentioned, GCN's goals include technological innovation as well as social impact. GCN assumes connectivity in rural areas will remain an ongoing challenge, while ML relies on the assumption that, with time, OneTouch coverage will be able to reach even the most remote doctors. This question echoes one of the fundamental tensions in ICTD research: as pragmatists, we aspire towards demonstrating the greatest social impact in real communities today, while as researchers, we try to identify what fundamental limitations exist and how these can be tackled in years to come. It is to be expected that untested technology would experience more hurdles in the short run, while its long-term contribution is yet to be seen. In short, then, a project can easily have a number of goals that might be frustrated by a lower-specificity solution.

A final note about lower-specificity solutions concerns their potential to be *too* widely appropriated. In the smallest district hospital (where GCN experienced greater adoption), both doctors tended to switch off or mute their phones while at the hospital in order to minimize distraction. Even in the larger urban hospitals, half of the 20 doctors who discussed their mobile phone usage claimed to keep their phones off while working because of distraction. Hospital administrators have related concerns; at one teaching hospital, a memorandum was circulated to all the doctors banning the use of mobile phones in many locations

*…to forestall the negative impact of mobile phone frequencies on medical equipments and improve the work ethics of staff. It is also announced… that it is a serious offence to disconnect life-supporting equipment in order to use their power sources (socket) to charge mobile phones.*

In view of concerns such as these, we suggest that researchers keep in mind both the (immediate) benefits and (eventual) costs of rapid adoption.

## VII. RELATED WORK

### ICTDs for Healthcare in the Developing World

The research on ICTDs to promote healthcare has a long and rich history [15],[34]. In the context of developing regions, remote medical consultation has been a popular and relatively successful approach [32]. Computers [3], mobile phones [25] and a combination of both [27] have demonstrated utility in a variety of settings.

### Framing the Problem

[21] provides a useful overview and categorization of how ICTs are conceptualized from a variety of different fields. Theory on 'framing the problem' can be drawn from social-constructivism, although we are not advocating here for a change from ICTD's traditional focus on pragmatism and advocacy/participatory research. Much of the work on qualitative methodology emphasizes the importance of reflexivity and self-awareness in order to minimize such bias [8], but the fact remains there are institutional accountabilities which no amount of methodology can shake. We can learn from the example of anthropologists who have challenged their own role in the colonial apparatus [11],[30]. For example, [17] provides an interesting comparison of how different sectors tackle the management of kiosks in India.

### Partnerships and Deployment

ICTD researchers have an important role to play at the crossroads of business administration and government policy, identifying how both contribute to the goal of development and how their contributions interplay. Many examples in ICTD research and development literature study the role of institutional players, although these papers are typically evaluative rather than action-oriented [18],[23],[24]. The literature on partnership selection and cultivation in ICTD is small but growing rapidly [26],[28],[31].

### Specificity

In reaction to the variable success of many deployments of general ICT, such as a number of telecenter and 'hole-in-the-wall' computing initiatives, much of the focus of the HCI ICTD research community has turned towards purpose-built technology [4],[33]. Many of those doing systems research focus instead on providing basic computing or connectivity functionality [13],[29]. [6] presents a useful survey of ICTD literature which mentions a pre-1999 trend towards non-specific technologies such as connectivity and a post-1999 trend towards specific solutions such as software and VOIP.

## VIII. CONCLUSION

In this paper, we have presented parallel case studies of two ICT projects addressing the need for improved medical consultation among doctors in Ghana. We have examined how the two projects have been shaped by the institutional context and the identity of their researchers. We looked at how the partnerships formed affect the solution outcome, delivery, and adoption; we argued that even the task of assessing technological infrastructure is far from objective; and we noted the implications of the 'specificity' of each solution.

We conclude with an open question raised by these discussions, summarized in Table 3. In ICTD research, much attention has been given to the socio-cultural, political, and economic contexts of target communities – yet 'difference' is a measure *between* communities; it is only by critically examining our own research community that we can understand the influence and impact of communities on each other. ICTD is linked with economic, social, political, and human development agendas [6]. Regardless of whether we take an integrated approach [10] or focus instead on specific local needs [14], there are still institutional, societal, and individual layers at play in our interventions. Researchers are embedded in contexts of existing friendships, collaborations, expertise and agendas, and we need to be conscious of what kind of consequences our decisions to draw on them have on project outcomes. In this paper we have seen impacts on everything from research agendas to infrastructure assumptions, yet the literature on these avenues of choice remain fragmented across a variety of other communities including CSCW and organizational behavior. ICTD researchers need to increase the exchange of ideas between these communities. For example, in the course of establishing GCN, the task of generating a viable 'business model' for the technology providers was not addressed, but a social entrepreneurship community would never allow such an omission. ML has had great adoption success, but technical infrastructure for evaluating its impact was neglected. We need a theoretical framework under which to unite the different fields of research for ICTD practitioners, one surrounding not just technology *design*, but also technology *framing, partnership, assumptions,* and *deployment*.

Table 3 – Impacts of project context on project decisions and outcomes

| Framing | By focusing on connectivity in the rural North, GCN targets a minority of doctors with the greatest need. |
| --- | --- |
| | By focusing on airtime cost, ML impacts the majority of doctors, the bulk of whom are in the urban South. |
| Partnerships and Deployment | Working through MOH allowed GCN to iterate rapidly with many hospitals early on, but with greater overhead later. GCN`s commitment to the U.S.-based GPSF was a key factor that tied it to a computer-based platform instead of mobile phones. |
| | GMA advertised ML effectively while OneTouch quickly took ownership of project execution and maintenance. ML's approach is highly dependent on the vagaries of a single national operator while GCN's is at the mercy of individual hospitals. |
| Infrastructure | GCN found the networks to be more challenged than anticipated. |
| | ML found mobile phone coverage lower than anticipated. |
| Solution | GCN`s higher specificity allowed it to incorporate evaluation indicators and address not only physical communication gaps but also gaps in social capital. Usage limited by access to computers in larger hospitals. |
| | ML adoption was much higher given the real-time nature of mobile phones and the fact that they were already widely tested and used. Lower specificity facilitated adoption and appropriation while making it difficult to evaluate the program. |